\documentclass[twocolumn,amsmath,aps]{revtex4}
\usepackage{graphicx,amssymb,upgreek}
\usepackage{hyperref}
\graphicspath{{Figures/}}
\begin{document}
%My commands
\newcommand{\be}{\begin{equation}}
\newcommand{\ee}{\end{equation}}
\newcommand{\bq}{\begin{eqnarray}}
\newcommand{\eq}{\end{eqnarray}}
\newcommand{\bsq}{\begin{subequations}}
\newcommand{\esq}{\end{subequations}}
\newcommand{\bc}{\begin{center}}
\newcommand{\ec}{\end{center}}
\newcommand {\R}{{\mathcal R}}
\newcommand{\al}{\alpha}
\newcommand\lsim{\mathrel{\rlap{\lower4pt\hbox{\hskip1pt$\sim$}}
    \raise1pt\hbox{$<$}}}
\newcommand\gsim{\mathrel{\rlap{\lower4pt\hbox{\hskip1pt$\sim$}}
    \raise1pt\hbox{$>$}}}

\title{Extended family of generalized Chaplygin gas models}

\author{V. M. C. Ferreira}
\email[Electronic address: ]{vasco.ferreira@astro.up.pt}
\affiliation{Instituto de Astrof\'{\i}sica e Ci\^encias do Espa{\c c}o, Universidade do Porto, CAUP, Rua das Estrelas, PT4150-762 Porto, Portugal}
\affiliation{Centro de Astrof\'{\i}sica da Universidade do Porto, Rua das Estrelas, PT4150-762 Porto, Portugal}
\affiliation{Departamento de F\'{\i}sica e Astronomia, Faculdade de Ci\^encias, Universidade do Porto, Rua do Campo Alegre 687, PT4169-007 Porto, Portugal}

\author{P. P. Avelino}
\email[Electronic address: ]{pedro.avelino@astro.up.pt}
\affiliation{Instituto de Astrof\'{\i}sica e Ci\^encias do Espa{\c c}o, Universidade do Porto, CAUP, Rua das Estrelas, PT4150-762 Porto, Portugal}
\affiliation{Centro de Astrof\'{\i}sica da Universidade do Porto, Rua das Estrelas, PT4150-762 Porto, Portugal}
\affiliation{Departamento de F\'{\i}sica e Astronomia, Faculdade de Ci\^encias, Universidade do Porto, Rua do Campo Alegre 687, PT4169-007 Porto, Portugal}

\date{\today}
\begin{abstract}
The generalized Chaplygin gas is usually defined as a barotropic perfect fluid with an equation of state $p=-A \rho^{-\alpha}$, where $\rho$ and $p$ are the proper energy density and pressure, respectively, and $A$ and $\alpha$ are positive real parameters. It has been extensively studied in the literature as a quartessence prototype unifying dark matter and dark energy. Here, we consider an extended family of generalized Chaplygin gas models parameterized by three positive real parameters $A$, $\alpha$ and $\beta$, which, for two specific choices of $\beta$ [$\beta=1$ and $\beta=\left(1+\alpha\right)/(2\alpha)$], is described by two different Lagrangians previously identified in the literature with the generalized Chaplygin gas.  We show that, for $\beta > 1/2$, the linear stability conditions and the maximum value of the sound speed $c_s$ are regulated solely by $\beta$, with $0 \le c_s \le 1$ if $\beta \ge 1$. We further demonstrate that in the non-relativistic regime the standard equation of state $p=-A \rho^{-\alpha}$ of the generalized Chaplygin gas is always recovered, while in the relativistic regime this is true only if $\beta=\left(1+\alpha\right)/(2\alpha)$. We present a regularization of the ($\alpha\rightarrow 0$, $A \rightarrow \infty$) limit of the generalized Chaplygin gas, showing that it leads to a logarithmic Chaplygin gas model with an equation of state of the form $p = {\mathcal A} \ln\left(\rho/\rho_{*}\right)$, where ${\mathcal A}$ is a real parameter and $\rho_*>0$ is an arbitrary energy density. We finally derive its Lagrangian formulation.

\end{abstract}
%\pacs{}
\maketitle 

\section{\label{sec:Introduction}Introduction}
Back in 1998 observations of distant supernovae of type Ia (SNIa) \cite{1998AJ....116.1009R,1999ApJ...517..565P} led to the discovery that the universe is currently expanding at an accelerated rate. Dark energy (DE) models can realize the observed late-time acceleration of the universe provided that the ratio between the proper pressure $p$ and energy density $\rho$ is sufficiently close to $-1$ near the present time ($p/\rho=-1$, in the simplest case of a cosmological constant $\Lambda$). In addition, a pressureless cold dark matter (CDM) component is required in order to explain the dynamics of large scale structures, such as galaxies and clusters of galaxies. In the standard cosmological model ($\Lambda$CDM model) about $95\%$ of the present energy density of the universe is composed by these two distinct dark components (CDM+DE), with most of the remaining $5\%$ being in the form of baryons. So far the $\Lambda$CDM model successfully explains a broad range of observational data, such as SNIa observations \cite{2012ApJ...746...85S}, baryonic acoustic oscillations (BAO) \cite{2014MNRAS.441...24A} or cosmic microwave background (CMB) anisotropies \cite{2016A&A...594A..13P}. However, there are outstanding fundamental questions regarding the nature of DM and DE \cite{2016PDU....12...56B} which motivate the search for extensions beyond the standard cosmological model, including dynamical DE, interacting DE or modified gravity (see \cite{2016arXiv160702979A} for a review).

Unified Dark Energy (UDE) models are also interesting in this regard, having the advantage of mimicking the DM and DE properties with a single underlying fluid. The first UDE prototype was the Chaplygin gas (CG) \cite{2001PhLB..511..265K}, usually defined as a perfect fluid with an equation of state given by $p=-A/\rho$ with $A>0$. The equation of state (EoS) parameter of the CG ($w\equiv p/\rho$) interpolates from $w=0$ at early times and $w=-1$ at late times when the CG energy density reaches a minimum value. This model attracted considerable interest also due to its connection to string theory d-branes \cite{2001PhLA..284..146B}. 

The CG model has been subsequently generalized to include an extra parameter $\alpha$ in the definition of its equation of state $p=-A/\rho^{\alpha}$ \cite{2002PhRvD..66d3507B}. For $0\leq\alpha\leq1$ the sound speed squared $c_{s}^{2} \equiv dp / d\rho=-\alpha w$ is greater than or equal zero (ensuring classical stability) and less than unity (ensuring that the sound speed is always subluminal). On the other hand, the case with $\alpha=0$ and finite $A$ is completely equivalent to the $\Lambda$CDM model \cite{2003JCAP...09..002A}. Furthermore, the generalized Chaplygin gas (GCG)  has been shown to be described by real scalar field Lagrangians \cite{2000physics..10042J,2007MNRAS.376.1169B,2006JMP....47c3101H,2007PhRvD..75b5008B} belonging to a sub-class of k-essence models \cite{2000PhRvD..62b3511C,2001PhRvD..63j3510A,2007APh....28..263D}. Several other k-essence models that can work as UDE have also been found in the literature (see, for example, \cite{2004PhRvL..93a1301S,2010AdAst2010E..78B}). 

The GCG has been claimed to be essentially ruled out \cite{2004PhRvD..69l3524S}, except for a very small region of parameter space very close to the $\alpha = 0$, due to the large absolute values of the linear sound speed attained at late times. However, it has been later realized that the impact of nonlinearities can dramatically change the evolution of the GCG (and UDE models in general) significantly enlarging the parameter space volume consistent with observations \cite{2004JCAP...11..008B,2014PhRvD..89j3004A,2014JCAP...10..036K}.\\

Here we investigate an extended family of GCG models, including two particular sub-classes previously identified in the literature with the GCG. The outline of the paper is as follows. In Sec. \ref{sec:Model} we introduce an extended family of GCG Lagrangians having a single extra parameter $\beta$ with respect to the standard GCG. We discuss its main features, in the relativistic and non-relativistic regimes, including the equation of state and sound speed of the underlying barotropic perfect fluid and their dependence on $\beta$. In this section we also determine sufficient conditions for linear stability and subluminal sound speeds. In Sec. \ref{sec:log} we investigate the properties of a logarithmic Chaplygin gas model obtained through the regularization of the $\alpha\rightarrow 0$, $A \to \infty$ limit of the GCG, and derive its Lagrangian formulation in the non-relativistic regime. We finally conclude in Sec. \ref{sec:conc}.

Throughout this paper we use units such that $c=1$, where $c$ is the value of the speed of light in vacuum, and we adopt the metric signature $(-,+,+,+)$. The Einstein summation convention will be used when a index variable appears twice in a single term, once in an upper (superscript) and once in a lower (subscript) position. Greek or latin indices are used for spacetime or spatial components, respectively.

\section{\label{sec:Model} Extended family of GCG models}

Consider a k-essence model with Lagrangian
\begin{equation}
\mathcal{L}\equiv\mathcal{L}\left(X\right)
\end{equation}
where $X = \frac{1}{2}\nabla_{\mu}\phi \nabla^{\mu}\phi$, $\nabla_{\mu}$ represents a covariant derivative with respect to the coordinate $x^{\mu}$, $\nabla^{\mu}=g^{\mu \nu} \nabla_{\nu}$, $g_{\mu\nu}$ are the components of the metric tensor, $g^{\mu \gamma} g_{\gamma \nu} = {\delta^\mu}_\nu$, ${\delta^\mu}_\nu$ is the Kronecker delta, and $\phi$ is a real scalar field. Provided that $\nabla_{\mu}\phi$ is timelike, the corresponding energy-momentum tensor defined by
\begin{equation}
T^{\mu\nu}=2\frac{\delta\mathcal{L}}{\delta g^{\mu\nu}}+g^{\mu\nu}\mathcal{L}\,,
\end{equation}
takes the form of a perfect fluid 
\begin{equation}
T^{\mu\nu}=\left(\rho+p\right)u^{\mu}u^{\nu}+pg^{\mu\nu}\,,
\end{equation}
with
\begin{eqnarray}
u_{\mu}=\frac{\nabla_{\mu}\phi}{\sqrt{2X}},  \qquad p=\mathcal{L}\left(X\right)\,, \qquad \rho=2Xp_{,X}-p\label{urhop}\,,
\end{eqnarray}
being, respectively, the components of the four-velocity, the proper pressure, and the proper energy density of the fluid (also $p_{,X}\equiv\partial p/\partial X$).

The k-essence Lagrangian
\begin{equation}\label{beta BI Lagrangian}
\mathcal{L}=-\rho_{\Lambda}\sqrt{\left(1-\left(2X\right)^{\beta}\right)^{\frac{2\alpha}{1+\alpha}}}\,,
\end{equation}
with $0\leq 2X\leq 1$, is a simple extension of the Lagrangians proposed in the literature to describe the GCG (with $\beta=1$ \cite{2000physics..10042J,2007PhRvD..75b5008B} and $\beta=\left(1+\alpha\right)/(2\alpha)$ \cite{2002PhLB..535...17B,2002PhRvD..66d3507B,2007MNRAS.376.1169B}). Here, $\alpha$ and $\beta$ are positive real model parameters, and $\rho_{\Lambda}$ is a positive constant energy density. Using Eq. \eqref{urhop}, the energy density 
\begin{equation}\label{rho_X}
\rho = \rho_{\Lambda}\left[1+\left(2X\right)^{\beta} \left(\beta\frac{2\alpha}{1+\alpha}-1\right)\right]\left(1-\left(2X\right)^ {\beta}\right)^{-\frac{1}{1+\alpha}}\,,
\end{equation}
can be computed for the proposed family of models described by Eq. \eqref{beta BI Lagrangian}. It is always non-negative for positive $\rho_{\Lambda}$, $\alpha$ and $\beta$. The corresponding equation of state parameter reads
\begin{equation}\label{EoS parameter}
w\equiv\frac{p}{\rho}=-\frac{1-\left(2X\right)^{\beta}} {1+\left(2X\right)^{\beta} \left(\beta\frac{2\alpha}{1+\alpha}-1\right)}\,,
\end{equation}
and is bounded between $-1$ and $0$ ($-1\leq w\leq 0$). The sound speed squared is given by
\begin{equation}\label{sound speed}
c_{s}^{2}\equiv\frac{p_{,X}}{\rho_{,X}} = \alpha\frac{1-\left(2X\right)^{\beta}}{1+\left(1+\alpha-\alpha\left(2X\right)^{\beta}\right)\left(\beta\frac{2\alpha}{1+\alpha}-1\right)}\,,
\end{equation}
with $c_{s}^{2}>0$ at all times if $\beta>1/2$. Given that, for $\beta>1/2$, $c_{s}^{2}\left(X\right)$ is a monotonically decreasing function of $X$, the maximum sound speed
\begin{equation}
c_{s,max}^{2}=\frac{1}{2\beta-1}\,,
\end{equation} 
is attained for $X=0$. On the other hand, the requirement that $c_{s}^{2}\leq 1$ at all times is satisfied if $\beta\geq 1$. Hence, the conditions of classical stability and subluminal sound speed are satisfied by this class of models if $\beta \ge 1$. Quantum stability is also ensured since the the Hamiltonian of the model described by Eq. \eqref{beta BI Lagrangian} is bounded from below (see for a discussion of the stability of k-essence models \cite{2006PhRvD..73f3522A}).\\

Rewriting Eq. \eqref{rho_X} as
\begin{equation}
\rho=\rho_{\Lambda}\left\lbrace 1+\left[1-\left(-\frac{p}{\rho_{\Lambda}}\right)^{\frac{1+\alpha}{\alpha}}\right]\left[\beta\frac{2\alpha}{1+\alpha}-1\right]\right\rbrace\left(-\frac{p}{\rho_{\Lambda}}\right)^{-\frac{1}{\alpha}}\,,
\end{equation}
it is clear that only the particular choice $\beta=\left(1+\alpha\right)/2\alpha$ leads to the standard GCG equation of state, i.e.
\begin{equation}\label{GCG EoS}
p=-\frac{A}{\rho^{\alpha}}\,,
\end{equation}
with $A=\rho_{\Lambda}^{1+\alpha}$, $w=-A\rho^{1+\alpha}$, and $c_s^{2}=-\alpha w$, independently of the value of $\rho$. 

\begin{flushleft}
\begin{figure}[!htbp]
\includegraphics[width=9cm]{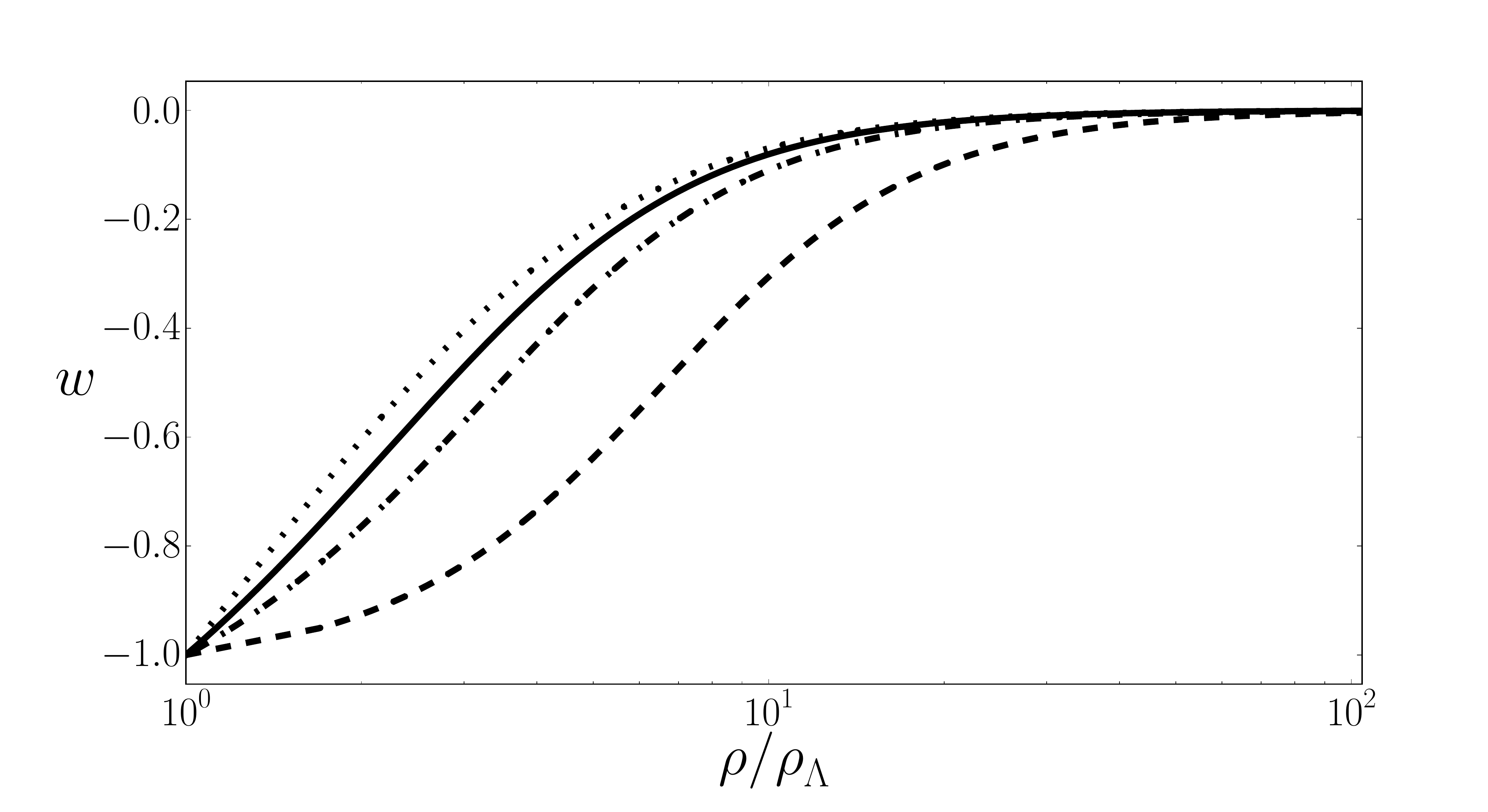}
\includegraphics[width=9cm]{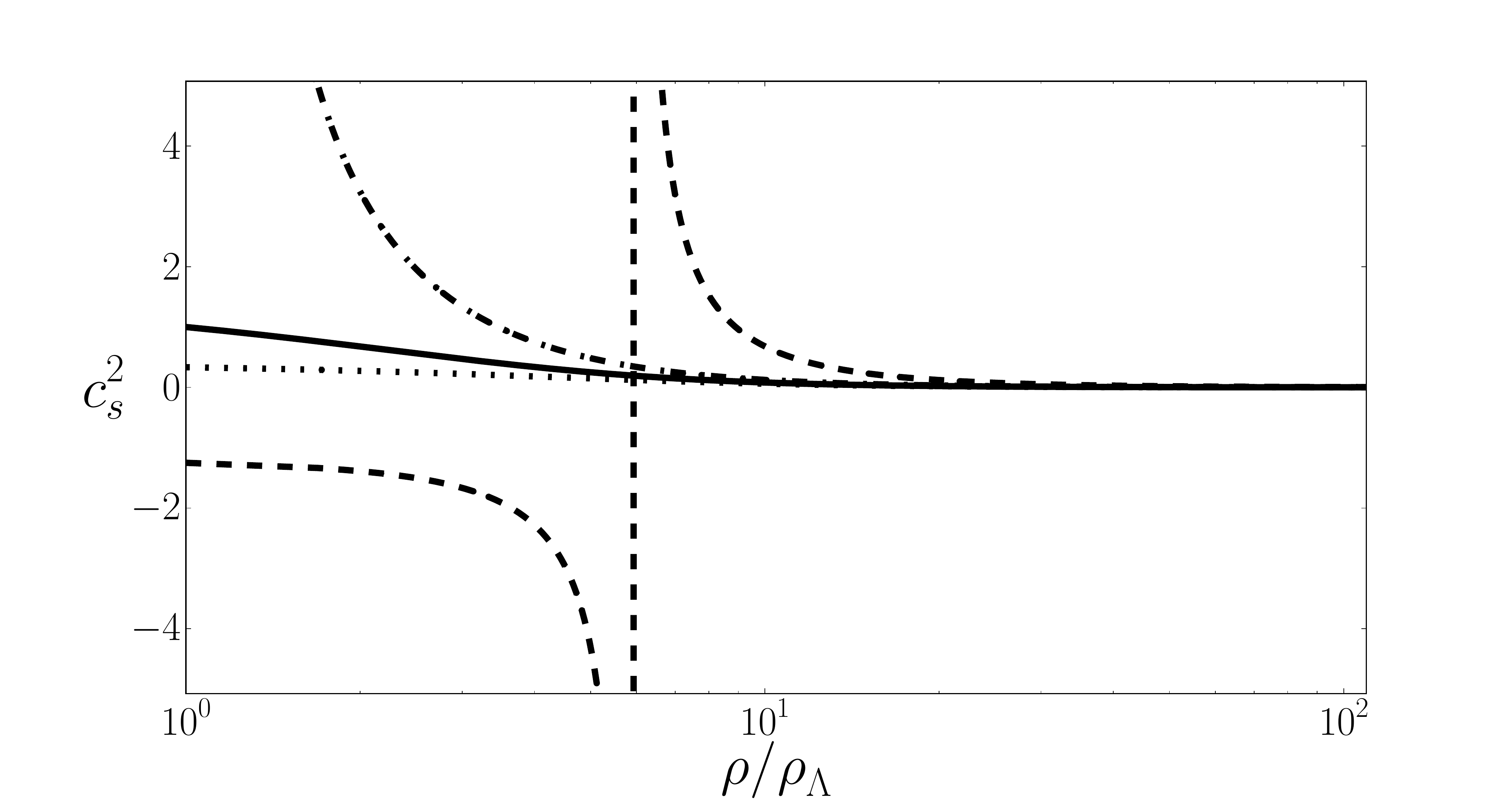}

\caption{The equation of state parameter $w$ and the sound speed squared $c_{s}^{2}$ as a function of $\rho$ for the GCG model ($\beta=\left(1+\alpha\right)/2\alpha$, solid line),($\beta=0.1$, dashed line), ($\beta=0.5$, dotted-dashed line) and ($\beta=2$, dotted line). Here, $\alpha$ has been fixed to unity.}\label{figure_w_cs2_X}
\end{figure}
\end{flushleft}

In Fig. \ref{figure_w_cs2_X} we plot EOS parameter $w$ (upper panel) and the sound speed squared $c_{s}^{2}$ (lower panel) as a function of the proper density $\rho$ for different values of $\beta$ [($\beta=\left(1+\alpha\right)/2\alpha$, solid line), ($\beta=0.1$, dashed line), ($\beta=0.5$, dotted-dashed line) and ($\beta=2$, dotted line)], while keeping $\alpha$ fixed to unity. The upper panel of Fig. \ref{figure_w_cs2_X} shows that the value of $w$ interpolates from $w=0$ ($\rho \gg \rho_\Lambda$) to $w=-1$ ($\rho =  \rho_\Lambda$) for any value of $\beta$ (this is also true for any other value of $\alpha > 0$). Nevertheless, it also shows that the shape of the function $p=p(\rho)$ is dependent on the value of $\beta$. The lower panel of Fig. \ref{figure_w_cs2_X} shows that for $\beta = 0.5$, $1$, and $2$ the sound speed $c_s$ is always a decreasing positive function of $\rho$, with $c_s \to 0$ for $\rho/\rho_\Lambda \to \infty$ and $c_s \to (2\beta-1)^{-1/2}$, for $\rho/\rho_\Lambda \to 1$ (note that this is true for any $\beta \ge 0.5$, and that if $\beta=0.5$ then $c_s \to \infty$ for $\rho/\rho_\Lambda \to 1$). On the other hand, it shows that for $\beta=0.1$ the sound speed $c_s$ is no longer a monotonic function of $\rho$ (in this case the sound speed diverges and changes sign at a specific value of $\rho > \rho_\Lambda$).
 
In the following we will show that the GCG equation of state is always recovered in the non-relativistic regime, independently of the value of $\beta$.

\subsection{Non-relativistic regime}\label{sec:GCG NR regime}

Consider the (high-density) non-relativistic regime of Eq. \eqref{beta BI Lagrangian} with $2X \sim 1$ and $\mathcal{L}_{NR}=p\approx 0$. The Taylor expansion of Eq. \eqref{beta BI Lagrangian} at first order in $\epsilon = 1 - 2X \approx 0$ gives
\begin{equation}\label{NR chaplygin gas}
\mathcal{L}_{NR}=-\rho_{\Lambda}\beta^{\frac{\alpha}{1+\alpha}}\sqrt{\left(1-2X\right)^{\frac{2\alpha}{1+\alpha}}}\,,
\end{equation}
which coincides with the Lagrangian considered in \cite{2000physics..10042J,2007PhRvD..75b5008B} to describe the generalized Chaplygin gas.

Using Eq. \eqref{urhop}, a relation between $p$ and $\rho$ may be computed explicitly:
\begin{equation}
\frac{\rho}{\rho_{\Lambda}}=\left(-\frac{p}{\rho_{\Lambda}}\right)\left(\frac{2\alpha}{1+\alpha}-1\right)+\beta\frac{2\alpha}{1+\alpha}\left(-\frac{p}{\rho_{\Lambda}}\right)^{-\frac{1}{\alpha}}\,.\label{rhoovrhoL}
\end{equation}
The first term on the right hand side of Eq. \eqref{rhoovrhoL} is negligible in the non relativistic regime ($p\approx 0$), thus leading to the standard GCG EoS, given by Eq. \eqref{GCG EoS}, with
\begin{equation}
A = \rho_{\Lambda}^{1+\alpha}\left(\frac{2\alpha}{1+\alpha}\beta\right)^{\alpha}\,.
\end{equation} 
Hence, the GCG equation of state is recovered in the non-relativistic regime for any $\beta>0$.

The usual non-relativistic Lagrangian of the GCG can be obtained by considering the irrotational non-relativistic dynamics of a barotropic perfect fluid \cite{2000physics..10042J,2007PhRvD..75b5008B}. Alternatively, it may be obtained from Eq. \eqref{NR chaplygin gas} with the identification $\phi=-t+\theta$, where $v_i = \theta_{,i}$ is the three-velocity of the fluid \cite{2007PhRvD..75b5008B}. Keeping only linear terms on $\dot \theta$
\begin{equation}\label{NR Lagrangian}
\mathcal{L}=-A^{\frac{1}{1+\alpha}}\left(2\beta\right)^{\frac{\alpha}{1+\alpha}}\sqrt{\left(\dot{\theta}+\frac{1}{2}\theta_{,i}\theta^ {,i}\right)^{\frac{2\alpha}{1+\alpha}}}\,,
\end{equation}
where a dot stands denotes a time derivative.

\section{\label{sec:log} Logarithmic Chaplygin gas}

In this section we shall consider a barotropic perfect fluid with sound speed squared given by
\begin{equation}\label{GCG_c2}
c_{s}^{2}\equiv\frac{dp}{d\rho}=\alpha\frac{A}{\rho^{1+\alpha}}\,,
\end{equation}
which is characteristic of the standard GCG [$\beta=(1+\alpha)/(2\alpha)$] independently of the value of $\rho$, and of the non-relativistic regime (large $\rho$) of the extended family of GCG models described by the Lagrangian given in Eq. \eqref{beta BI Lagrangian}. The corresponding EOS parameter for $\alpha \neq 0$,
\begin{equation}\label{GCG EOS}
w \equiv \frac{p}{\rho} =- \frac{A}{\rho^{1+\alpha}}  + \frac{C}{\rho} =-\frac{c_s^2}{\alpha} + C \left(\frac{c_s^2}{\alpha A}\right)^\frac{1}{1+\alpha} \,,
\end{equation}
may be obtained by integrating Eq. (\ref{GCG_c2}), where $C$ is a real constant which is usually assumed to be equal to zero.

Let us rewrite Eq. \eqref{GCG EoS} as
\begin{equation}
p=-\frac{A}{\rho_*^{\alpha}}\left(\frac{\rho_{*}}{\rho}\right)^{\alpha}+C\,,
\end{equation}
with $A/\rho_*^{\alpha}=\rho_{\Lambda}$. Expanding around $\alpha=0$ one obtains
\begin{equation}
p=A\left[-1 + \alpha\ln\left(\frac{\rho}{\rho_{*}}\right) + \mathcal{O}\left(\alpha^{2}\right)\right]+C\,.
\end{equation}

Consider the following limit
\begin{equation}\label{limit}
\mathcal{A}=\lim_{\substack{ A \to \infty  \\ \alpha\to 0}} \alpha A\,,
\end{equation}
with finite $\mathcal{A}$ and $C=A$. In this limit, the EoS parameter is given by
\begin{equation}\label{logCG}
w \equiv \frac{p}{\rho}=\frac{\mathcal{A}}{\rho}\ln\left(\frac{\rho}{\rho_{*}}\right)\,,
\end{equation}
and the sound speed squared is equal to
\begin{equation}\label{logCG_c2}
c_{s}^{2} \equiv \frac{dp}{d \rho} =\frac{\mathcal{A}}{\rho}\,.
\end{equation}
The model defined by Eq. \eqref{logCG} shall be referred as Logarithmic Chaplygin gas (logCG), and is one of the simplest extensions of the standard $\Lambda$CDM model containing a single extra parameter. By tuning the value of $\mathcal{A}$ and the reference density $\rho_{*}$ this model allows for the the study small deviations from the $\Lambda {\rm CDM}$ model. On the other hand, the $\Lambda$CDM model can be obtained by considering the $\mathcal{A} \to 0$, $\rho_* \to \infty$ limit, with finite $\mathcal{A} \ln \rho_*$. 

\subsection{Relativistic Lagrangian formulation}

The relativistic form ${\mathcal L}(X)$ of the Lagrangian of the logCG may be found taking into account that if $p={\mathcal L} = {\mathcal A} \ln (\rho/\rho_*)$ then 
$\rho=2X p_{,X} -p = \rho_* e^{p/A}$, which implies that
\begin{equation}
\frac{dX}{X}=2 \frac{dp}{p+\rho_* e^{p/A}}\,
\end{equation}
or, equivalently,
\begin{equation}
\ln \left(\frac{X}{X_*}\right) = 2 \int_{p_*}^p \frac{dp'}{p'+\rho_* e^{p'/A}}\, \label{LXlog}
\end{equation}
where $X_*$ is an arbitrary integration constant. Unfortunately, the right hand side of Eq. \eqref{LXlog} does not have a simple analytical solution, and has to be evaluated numerically. Nevertheless, in the following we shall obtain an analytical form of the Lagrangian valid in the non-relativistic regime.

\subsection{Lagrangian formulation: non-relativistic regime}

Here we shall derive the non-relativistic Lagrangian for the logCG model obeying an EoS given by Eq. \eqref{logCG}. In classical fluid dynamics the Hamiltonian of an irrotational perfect fluid is given by
\begin{equation}
H\left(\rho,\theta,t\right)=\int dx^{3}\mathcal{H}=\int dx^{3}\left(\frac{1}{2}\rho \, \theta_{,i}\theta^{,i}+V\left(\rho\right)\right)\,,
\end{equation}
where $V\left(\rho\right)$ is some potential, and $\mathcal{H}\left(\rho,\theta,x^{i},t\right)$ is equal to
\begin{equation}
\mathcal{H}\left(\rho,\theta,x^{i},t\right)=\dot{\rho} \, \theta - \mathcal{L}\left(\rho,\rho_{,i},\dot{\rho},x^{i},t\right)\,,
\end{equation}
The Lagrangian reads
\begin{equation}\label{Lagrangian fluid dynamics}
\mathcal{L}=\dot{\rho}\theta-\frac{1}{2}\rho \, \theta_{,i}\theta^{,i}-V\left(\rho\right)\,,
\end{equation}
with $\rho$ and $\theta$ being canonically conjugate, i.e.
\begin{equation}\label{phi}
\theta=\frac{\partial\mathcal{L}}{\partial\dot{\rho}}\,,
\end{equation}
and
\begin{equation}\label{phi dot}
\dot{\theta}=\frac{\partial\mathcal{L}}{\partial\rho}=-\frac{1}{2}\theta_{,i}\theta^{,i}-\frac{dV}{d\rho}\,.
\end{equation}
Assuming isentropic motion then
\begin{equation}
p=\rho\frac{dV}{d\rho}-V\,, \label{pV}
\end{equation}
where the enthalpy is equal to $dV/d\rho$. Eqs. \eqref{logCG} and \eqref{pV} imply that 
\begin{eqnarray}
V&=&-\mathcal{A}\left[1+\ln\left(\frac{\rho}{\rho_{*}}\right)\right]+D \rho \\ 
&=& -\mathcal{A}\left[1+\ln\left(\frac{\rho}{\rho_{*}}\right)\right] \,,
\end{eqnarray}
where $D$ is an arbitrary integration constant which is taken to be zero. Using  Eq. \eqref{phi dot} one finds
\begin{equation}
\rho=\mathcal{A}\left(\dot{\theta}+\frac{1}{2}\theta_{,i}\theta^{,i}\right)^{-1}\,.
\end{equation}

One may now eliminate $\rho$ from Eq. \eqref{Lagrangian fluid dynamics} to obtain
\begin{equation}\label{Lagrangian logCG}
\mathcal{L}=\mathcal{A}\left[\ln\left(\frac{\mathcal{A}}{\rho_{*}}\right)- \ln\left(\dot{\theta}+\frac{1}{2}\theta_{,i}\theta^{,i}\right)\right]\,.
\end{equation}

Eq. \eqref{Lagrangian logCG} may also be obtained by considering the lagrangian given in Eq. \eqref{NR Lagrangian} 
\begin{equation}\label{NR Lagrangian w/const.}
\mathcal{L}=-A^{\frac{1}{1+\alpha}}\left(\frac{1+\alpha}{\alpha}\right)^{\frac{\alpha}{1+\alpha}}\sqrt{\left(\dot{\theta}+\frac{1}{2}\theta_{,i}\theta^ {,i}\right)^{\frac{2\alpha}{1+\alpha}}}+A\,,
\end{equation}
with an added constant term $A$ which does not affect the dynamics of $\theta$, and $\beta=(1+\alpha)/(2\alpha)$. 

Performing a Puiseux series expansion of Eq. \eqref{NR Lagrangian w/const.} around $\alpha=0$ and keeping up to first order terms in $\alpha$ one obtains
\begin{equation}\label{Lagrangian after expansion}
\mathcal{L}=\alpha A\left[\ln\left(\alpha\frac{A}{\rho_{*}}\right)-\ln\left(\dot{\theta} +\frac{1}{2}\theta_{,i}\theta^{,i}\right)\right]\,.
\end{equation}
Considering the $\alpha \to 0$, $A \to \infty$ limit with finite $\mathcal{A}=\alpha A$, as given by Eq. \eqref{limit}, then the Lagrangian given in Eq. \eqref{Lagrangian after expansion} is equal to that given in Eq. \eqref{Lagrangian logCG}. It describes a non-relativistic perfect fluid with a logCG EoS.

\section{\label{sec:conc} Conclusions}
In this paper we presented a Lagrangian formulation of an extended family of perfect fluid models which include two particular sub-classes identified in the literature with the GCG. We have shown that in the non-relativistic (high density) regime these models are characterized by the GCG standard equation of state, and asymptotically approach a cosmological constant in the relativistic (low density) regime. This extension of the standard GCG model includes a single parameter $\beta$, which we have shown to control both the linear stability conditions and the maximum value of the sound speed, the standard GCG model being recovered for $\beta=\left(1+\alpha\right)/(2\alpha)$. We have also demonstrated that a regularization of the ($\alpha \to 0$, $A \to \infty$) limit of the GCG gives rise to a logCG model, which may be regarded as the simplest one parameter extension of the standard $\Lambda$CDM model. \\
%%%%%%%%%%%%%%%%%%%%%%%%%%%%%%%%%%%%%%%%%%%%%%%%%%%%%
\begin{acknowledgments}

V.M.C. Ferreira was supported by the FCT fellowship (PD/BD/135229/2017), within the FCT PD Program PhD::SPACE (PD/00040/2012). Funding of this work has also been provided by the FCT grant UID/FIS/04434/2013. This paper benefited from the participation of the authors on the COST action CA15117 (CANTATA), supported by COST (European Cooperation in Science and Technology).

\end{acknowledgments}

%%%%%%%%%%%%%%%%%%%%%%%%%%%%%%%%%%%%%%%%%%%%%%%%%%%%%%%%%%

\bibliography{Born_Infeld_generalization}

\end{document}